\documentclass[prl,twocolumn,superscriptaddress,showpacs]{revtex4-1}

\usepackage[dvips]{graphicx}
\usepackage{amsmath}
\usepackage{latexsym}

\begin{document}
\title{Phyllotactic description of hard sphere packing in cylindrical channels}
\date{\today}

\author{A. Mughal}
\affiliation{Institute of Mathematics and Physics, Aberystwyth University, Penglais, Aberystwyth, Ceredigion, Wales, SY23 3BZ}

\author{H. K. Chan} 
\affiliation{Foams and Complex Systems, School of Physics, Trinity College Dublin, Dublin 2, Ireland}
\affiliation{Department of Applied Physics, Hong Kong Polytechnic University, Hong Kong}

\author{D.  Weaire} 
\affiliation{Foams and Complex Systems, School of Physics, Trinity College Dublin, Dublin 2, Ireland}

\begin{abstract}
We develop a simple analytical theory that relates dense sphere packings in a cylinder to corresponding disk packings on its surface. It applies for ratios $R=D/d$ (where $d$ and $D$ are the diameters of the hard spheres and the bounding cylinder, respectively) up to $R=1+1/\sin(\pi/5)$. Within this range the densest packings are such that all spheres are in contact with the cylindrical boundary. The detailed results elucidate extensive numerical simulations by ourselves and others by identifying the nature of various competing phases.

\end{abstract}

\pacs{64.75.Yz, 87.18.Hf, 81.16.Rf}

\maketitle


Packing problems, such as that of finding the densest arrangement of hard spheres, are among the most ancient and challenging problems in mathematics and science. The most famous example is the Kepler conjecture (1611), concerned with identical spheres in an unbounded volume \cite{Hales:1992}. In contrast, the optimal packing of such spheres within a {\it bounded} space is much less well understood. A prominent example is finding the densest packing arrangement of monodisperse spheres within a cylindrical channel, which we consider here. There are extensive results from simulations, which we corroborate and augment in this Letter. We supply a detailed interpretation of them, by making a connection with the analogous (but simpler) problem of packing disks on the bounding surface, which is an interesting problem in its own right \cite{Bleicher:1964p865}. 

Our procedure is therefore to first explore an analytical treatment of the surface disk packing problem, and then develop a transformation which gives approximate but illuminating results for sphere packing within a cylinder. Such surface patterns have their origin in the early study of intriguing spiral arrangements of leaves and stems \cite{Airy:1872}. Thus they are associated with the term {\it phyllotaxis} which now stands for the standard classification of triangular patterns on a cylinder, in any context. It is defined and used, in what follows, together with the corresponding notation for rhombic patterns.

Both packing problems find numerous counterparts in nature and in the laboratory; examples include biological microstructure \cite{Erickson:1973}, the morphology of dry \cite{Hutzler:2002} and wet foams \cite{Chan:2010} in tubes, colloids in microchannels \cite{Tymczenko:2008}, and the packing of fullerenes in nanotubes \cite{Khlobystov:2004}. 

Simulations intended to search for the densest cylindrical arrangement of spheres  \cite{Pickett:2000} have revealed a sequence of helical columnar phases, as the ratio $R=D/d$ is varied (where $d$ and $D$ are the diameters of the hard spheres and the bounding cylinder, respectively). Some of these structures have been recognized in various experimental contexts, but there has until now been no clear understanding of what determines the sequence of structural transitions as the cylinder diameter is increased. In this letter we demonstrate that the key to this progression is the familiar  phyllotactic description. Using a straightforward and elementary method we provide a full taxonomy of competing phases and also elucidate the geometric relationships between them. Surprisingly, we can use this method to also provide a full semiquantitative description of the helical structures together with their packing densities. In general the densest structures are of the kind called ``staggered'' by Pickett \cite{Pickett:2000}: we shall use the term ``line slip''.

\begin{figure}
\begin{center}
\includegraphics[width=0.95\columnwidth ]{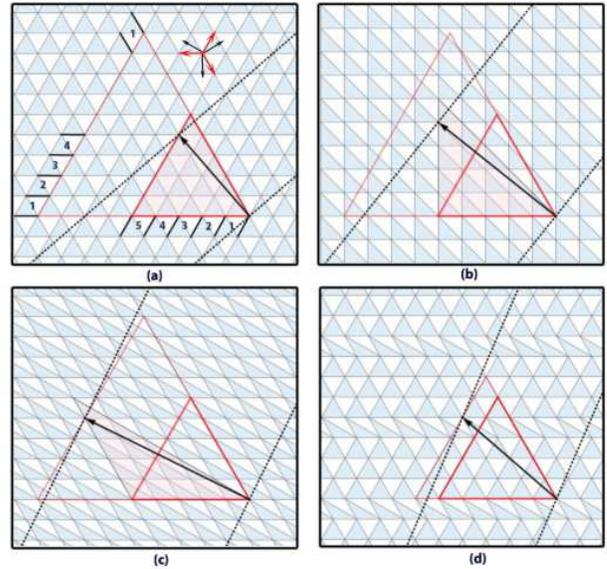}
\caption{(Color online). (a) Symmetric packing $[5,4,1]$ - the black and red arrows indicate the  periodicity ${\bf V}$ and primitive lattice vectors, respectively; (b) square packing $[5,4]$; (c) symmetric packing $[9,5,4]$ which is connected to the structure $[5,4,1]$ by an affine shear; (d) a line-slip structure between the symmetric packings  [6,4,2] and [5,4,1].}
\label{distortion_operator}
\end{center}
\end{figure}

We begin by discussing the disk packing problem. It is known that the densest packing arrangement of disks on a plane is that of a symmetric triangular lattice with lattice spacing $d'$ (where $d'$ is the disk diameter) \cite{Toth:1953}. A piece of the lattice can be excised and  wrapped onto a cylinder of diameter $D'$  (without any mismatch at the wrap edges) if there exist two lattice points separated by a vector ${\bf V}$ of magnitude,
\begin{equation}
|{\bf V}|=\pi D'.
\label{eq:magnitude_of_V}
\end{equation}
Thus  ${\bf V}$ plays the role of the periodicity vector, as shown in Fig (\ref{distortion_operator}a) (where the lattice vertices correspond to the disk centers). The excised section is bounded by two parallel lines perpendicular to ${\bf V}$, these are the wrap edges and correspond to lines with  $\theta=0$ and $\theta=2\pi$, respectively, on the surface of a cylinder described by polar coordinates $\left(\rho=\frac{D'}{2}, \theta, z \right)$. 

In the phyllotactic scheme the periodicity vector ${\bf V}$ can be defined by a set of three, ordered, positive integers $(l=m+n,m,n)$, known as parastichy numbers in biology. Any pair of these numbers are the absolute values of the coefficients expressing ${\bf V}$ as a combination of two basis vectors, chosen from among three possibilities.  Fig (\ref{distortion_operator}a) shows an example for the structure $[l,m,n]=[5,4,1]$ along with the basis vectors of the triangular lattice.

The resulting structure, when wrapped onto a cylinder, is a dense, homogenous packing - in the sense that all disk sites are equivalent - which we call a {\it symmetric} lattice packing. Between the discrete values of $D'$, and fixed $d'$, for which such symmetric packings exist we first consider the most obvious adjustment to fit the cylinder. In this the required periodicity is accommodated by an affine transformation, or strain, of the lattice (more complex arrangements are possible but we do not find it necessary to consider them). The affine transformation distorts the equilateral triangle into an isosceles triangle by keeping the length of two adjacent sides fixed and varying the third. The imposed strain has three free parameters; these can be chosen to satisfy Eq. ($\!\!$~\ref{eq:magnitude_of_V}) and give each disk four contacting neighbors (in any one of three ways). 

As $D'$ is varied, the strained structure proceeds from one symmetric packing to another, as shown in Fig (\ref{distortion_operator}c). The rules for this process, when applied to the second and third phyllotactic indices, are as follows \cite{Mughal}:
\begin{eqnarray}
(m,n)
&\rightarrow&
(m-n, n)
\nonumber
\\
(m,n)
&\rightarrow&
(m+n, m)
\nonumber
\\
(m,n)
&\rightarrow&
(m+n, n),
\nonumber
\end{eqnarray}
where the new indices may have to be rearranged into descending order. For example, in the case of the symmetric packing $[5,3,2]$ the above rules yield $(3,2)\rightarrow(2, 1)$, $(3,2)\rightarrow(5, 3)$ and $(3,2)\rightarrow(5, 2)$, where a rearrangement of the indices in the first case was necessary. Thus the symmetric packing $[5,3,2]$ is connected, by an affine shear, to the symmetric packings $[3,2,1]$, $[8,5,3]$, and $[7,5,2]$. 

The surface density (or area fraction of the disks on the plane) has a maximum value for a symmetric structure. Between any two symmetric structures, connected in the way described above, the surface density has a lower value since each disk now has only four contacting neighbors. Such intermediate arrangements can be labeled by using rhombic notation $[p,q]$ (see \cite{Erickson:1973}) and have a minimum surface density when the disks are arranged into a square lattice. Figure (\ref{distortion_operator}c) shows such a square packing intermediate between two symmetric packings.

We shall call these {\it asymmetric} lattice packings. In general, they are superseded by another type, as follows (indeed, one can show by convexity arguments that the asymmetric lattice structure is always unstable with respect to perturbations, under pressure).

The second type of packing is achieved by an inhomogeneous deformation of the symmetric lattice in which there is a localized strain or slip along a line (and its periodic replicas) as shown in Fig (\ref{distortion_operator}d). Again we have three choices of the direction of slip. The rest of the structure remains symmetric and close-packed, i.e. with six contacts for every disk. This \textit{line-slip} distortion, as we shall call it, Êcan again be expressed analytically. ÊOther possibilities, involving less localized strain exist, but these yield a lower density. Hence we include only the (optimized) case of single \textit{line-slip solutions}. 

Again there is a simple rule for the close-packed structures that are the endpoints of line-slip solutions; in view of their importance, we will identify explicitly the three types of solution \cite{Mughal}:
\begin{eqnarray}
(1)  \;\;\;\;\;\;\;\;\; (m,n)
&\rightarrow&
(m+1, n) \;\;\;\;\;\;\;\;\;  \textrm{or} \;\;\; (m-1, n) 
\nonumber
\\
(2)  \;\;\;\;\;\;\;\;\; (m,n)
&\rightarrow&
(m, n+1) \;\;\;\;\;\;\;\;\;  \textrm{or} \;\;\; (m, n-1) 
\nonumber
\\
(3)  \;\;\;\;\;\;\;\;\; (m,n)
&\rightarrow&
(m+1, n-1) \;\;\; \textrm{or} \;\;\; (m-1, n+1), 
\nonumber
\end{eqnarray}
where the leading numbers denote the direction $\widehat{\bf a}_1$,  $\widehat{\bf a}_2$ or $\widehat{\bf a}_3$ of the line-slip. Again the above rules apply to the second and third phyllotactic indices of a given close-packed structure and keep either $n$, $m$ or $l$ constant. For example, by using the above rules, the symmetric packing $[5,3,2]$ is connected by a line slip along $\widehat{\bf a}_1$ to $[6,4,2]$ or $[4,2,2]$, a line slip along $\widehat{\bf a}_2$ yields $[4,3,1]$ or $[6,3,3]$ and a line slip along $\widehat{\bf a}_3$ yields $[5,4,1]$ or $[5,3,2]$ (note in the second case along $\widehat{\bf a}_3$  that a rearrangement of the phyllotactic indices into descending order was necessary, the new structure has the same indices as the initial one). Degeneracy is also possible; e.g. $[4,2,2]$ is connected to $[5,3,2]$ by a line slip along $\widehat{\bf a}_1$ or $\widehat{\bf a}_2$.

\begin{figure}
\begin{center}
\includegraphics[width=1.0\columnwidth ]{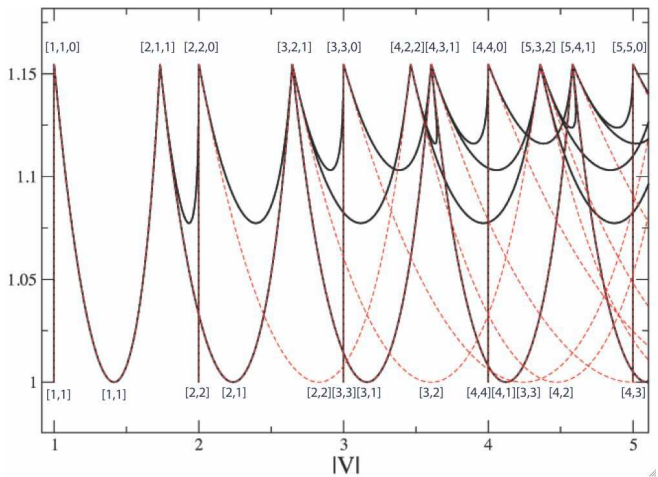}
\caption{(Color online). Area fraction of disk packings on the plane which are consistent with wrapping onto a cylinder of diameter $D'=|V|/\pi$. Asymmetric packings are shown by (red) dashed curves; line-slip solutions are (black) continuous curves. }
\label{planar_states}
\end{center}
\end{figure}

Such solutions (inspired by our simulations - see below) are the densest we have found by analytical searches, but we can offer no proof that excludes other denser structures of greater complexity. Fig (\ref{planar_states}) presents the analytic results (to be published in detail elsewhere    \cite{Mughal}) for the surface density of all of the packings described above, where we have set $d'=1$. Asymmetric packings are shown by (red) dashed curves, line-slip solutions by (black) continuous curves; the symmetric packings are the labeled peaks with maximum density, while the lines connecting them are labeled by using the rhombic notation. 

\begin{figure}
\begin{center}
\includegraphics[width=1.0\columnwidth ]{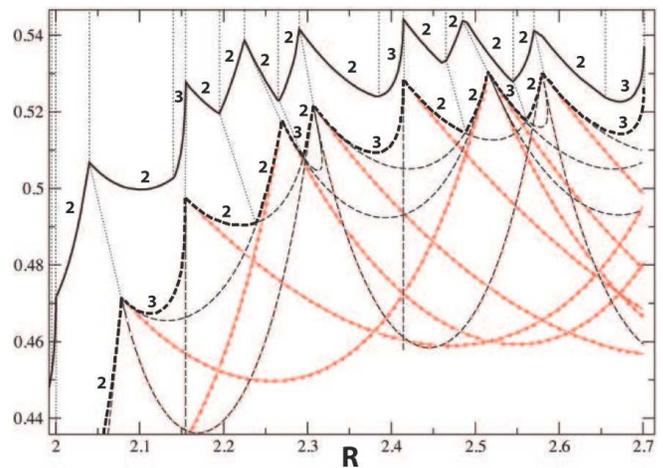}
\caption{(Color online). Comparison of simulated and analytic volume fractions for the densest packings, upper and lower curves respectively. Dotted lines are a guide for the eye to the detailed correspondence. In the analytic results, the line-slip structures are identified by black dashed lines and the red dashed-dotted lines denote asymmetric packing structures. The numbers 1,2 and 3 (see text) denote the type of line slip observed (upper curve) or predicted (lower curve) - with $1 \backslash 2$ denoting a degenerate case.}
\label{vf_results}
\end{center}
\end{figure}

These results already show a strong resemblance to the corresponding simulation data for sphere packings in cylinders (but before we make a final comparison we will make a further transformation). The heavy continuous (black) line in  Fig (\ref{vf_results}) indicates the computed optimal packing density (or volume fraction) of monodisperse hard spheres in an infinitely long cylindrical tube for $R(=D/d)$. The present analysis was stimulated by that  the numerical work of Pickett et al \cite{Pickett:2000}, who used the method of simulated annealing to investigate the cylindrical packing of hard spheres. We have employed two alternative approaches to verify and extend their findings. One is based on simulated annealing and employs the use of spiral boundary conditions to make the search for optimal structures more efficient, the other on sequential deposition. Both will be described elsewhere. The data of Fig (\ref{vf_results}) is obtained by the method with spiral boundary conditions, but the results of all three approaches are consistent. 

\begin{figure}
\begin{center}
\includegraphics[width=0.85\columnwidth ]{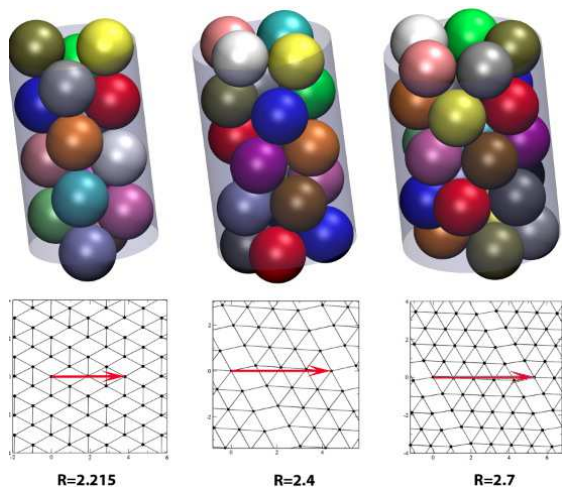}
\caption{(Color online). Simulated 3D structures for various values of R, with the corresponding ``rolled out'' phyllotactic diagram below (the red arrow denotes the periodicity vector).}
\label{sphere_packing_states}
\end{center}
\end{figure}

A representative sample of the simulated 3D structures is shown in Fig (\ref{sphere_packing_states}). Below each structure is the ``rolled out''  phyllotactic pattern.The sphere centers lie on an inner cylinder $D'=D-d$ and can be mapped to the plane by using Cartesian coordinates $\left[(D' \theta)/ 2, z \right]$, where they are shown as black dots. Contacting spheres are indicated by dots connected by black lines. The periodicity vector, between two identical points on the cylinder, is denoted by a red arrow. The relationship to planar line-slip structures is evident.

We relate the disk packings to those of spheres in contact with a cylinder, such as those shown in Fig (\ref{sphere_packing_states}), as follows. Recall that the sphere centers lie on an inner cylinder of diameter $D'$. ÊThe sphere packing density is trivially related to the point density of centers on that cylinder and, hence, to that of the planar pattern of points obtained by rolling out the cylinder on a plane. The qualitative resemblance to the disk problem becomes obvious, but its exact quantitative expression is not straightforward. Two sphere centers, separated by a vector $\textbf{r}$ in 3D correspond to points in 2D whose separation $S$ depends on the orientation of Ê$\textbf{r}$. For $D'>d$, the two extreme cases are $S = |\textbf{r}|$, for $\textbf{r}$ parallel to the cylinder axis, and $S = D'\textrm{sin}^{ - 1} (|\textbf{r}|/D')$, for $\textbf{r}$ perpendicular to the cylinder axis. We adopt, as an ansatz, an interpolation between these two limits which has the correct analytic form, 
\begin{equation}
S^2 = Ê\left[|\textbf{r}| \sin(\phi)\right]^2 + ÊÊÊ\left[D' Ê\sin^{-1} (|\textbf{r}|/D') \cos(\phi)\right]^2,
\label{eq:transformation}
\end{equation}
where $\phi$ is the angle that the points on the inner cylinder make with the vector ${\bf V}$. This rule is applicable for $R \geq 2$ and a  2D packing that is consistent with it is simply a packing of identical ellipses, whose major axes are all parallel to the vector ${\bf V}$ defined above, and may be obtained from any one of the disk packings by a simple stretch in the direction of ${\bf V}$ (which also changes ${\bf V}$ itself). In this way we arrive at a transformation of the circular disk packings into an approximate representation of corresponding sphere packings, with densities as shown in ÊFig (\ref{vf_results}). 

This is a good approximation for our purposes; we note that the ratio between the predicted volume fraction, using our method, and numerical results is already 0.9426 at $R=2.15$ (i.e. the symmetric [3,3,0] structure) and continues to rapidly improve as $|{\bf V}|$ becomes larger. A more cumbersome method that uses the exact dependence of $dÕ$ on $\phi$ is perfectly possible, but a small sacrifice of accuracy preserves the extreme simplicity and transparency of our approach. By reference to the corresponding disk packing the approximate description developed above identifies the nature of all of the simulated sphere packing structures involved and is mostly correct in picking the optimal structure. Transition between different sphere packings structures, indicated by the vertical dotted lines at the top of Fig (\ref{vf_results}), can be understood by reference to the analogous disk packing problem.

There are two situations leading to qualitative discrepancies between the theory and numerical results. The first corresponds to the region $R\leq2.0$ for which the approximation Eq. ($\!\!$~\ref{eq:transformation}) fails. The result is that the twisted zigzag structure \cite{Pickett:2000} arises unexpectedly as a \textit{stable} asymmetric lattice structure. The transition to the line-slip structure at ÊR Ê= 1.995 is therefore of a different character from the others. Above this point, all of the intermediate Êstructures are derived from the planar line-slip configurations. They are of the ÒstaggeredÓ spiral Êcharacter  (see the second and third figures in Fig (\ref{sphere_packing_states})), noted by Pickett et al in their simulations \cite{Pickett:2000},  and exist up to $R=1+1/\sin(\pi/5)$. Beyond this point not all the spheres are in contact with the boundary and the correspondence with the phyllotactic scheme is lost, although it may prove useful in the analysis of the surface layers. 

The second discrepancy is due to neglecting higher order corrections to $S$ in Eq. ($\!\!$~\ref{eq:transformation}) and leads to the observation of line-slip solutions of type $(2)$, in the numerical data, where type $(3)$ would have been expected. This effect, whereby type $(2)$ solutions have a greater than expected volume fraction, diminishes for larger  $|{\bf V}|$ as the higher order corrections to $S$ fade away. 

A notable detail that emerges in Êthe planar lattice analysis, and is retained in transformation to the sphere packing problem, is the singular behavior just before the points labeled 220, 330, etc. For reasons of symmetry, the decrease of density takes a square-root form and this is observed in the simulation data (most clearly in plots of the derivative of the density with respect to Ê$D$).

In summary, for $2\leq R \leq1+1/\sin(\pi/5)$, each phyllotactic state (l,m,n) corresponds to a local density maximum. Because of periodic constraints, and the nonoverlapping nature of hard spheres, these maxima are attained only at discrete values of $D$ where six contacts are possible. Where only four contacts are possible, the line-slip structure with the highest density is observed. The density of the line-slip structure varies continuously with $D$, but transitions between them involve a structural discontinuity. Though we are not aware of any experimental observations of line-slip structures, transitions between phyllotactic states have been observed in other systems \cite{Nisoli:2009}. In further work  we will provide full details of the analytical results used here \cite{Mughal} and include questions of stability and chirality. We shall extend our numerical results to larger cylinder diameters - the packing problem then changes to one in which not all spheres touch the cylinder. 
 
We thank S. Hutzler for useful discussions. DW \& HKC acknowledge the financial support of SFI  \& ESA.  DW thanks the University of Hyderabad for hospitality during the period in which this work was completed. 

\bibliographystyle{nonspacebib}

\end{document}